\documentclass[12pt]{article}
\usepackage[xdvi]{graphicx}
\usepackage{amssymb}
\newcommand{\bea}{\begin{eqnarray}}
\newcommand{\eea}{\end{eqnarray}}

\makeatletter
\@addtoreset{equation}{section}
\makeatother

\begin{document}
\begin{titlepage}
\begin{flushright}
OU-HET 635/2009
\end{flushright}

\vspace{25ex}

\begin{center}
{\Large\bf 
Simple mass relations for bulk fermions
}
\end{center}

\vspace{1ex}

\begin{center}
{\large 
Nobuhiro Uekusa
}
\end{center}
\begin{center}
{\it Department of Physics, 
Osaka University \\
Toyonaka, Osaka 560-0043
Japan} \\
\textit{E-mail}: uekusa@het.phys.sci.osaka-u.ac.jp
\end{center}


\vspace{3ex}

\begin{abstract}

Relations between bulk mass parameters
for fermions propagating in higher dimensions are
studied in analogy with the empirical mass relation for charged leptons.
Masses of three generation of four-dimensional charged leptons
are achieved from the bulk mass parameters of 
the same-order values. 
We find that
the observed pattern of charged lepton mass spectrum
is accommodated approximately in 
a set of 
simple relations for bulk and physical masses.

\end{abstract}
\end{titlepage}

\section{Introduction}

The hierarchical structure such as macroscopic
and microscopic pictures has provided 
an intuitive and
predictable way to
understand physics as effective field theory.
The type of    
dynamical variables to describe phenomena in the theory
depends on the energy scale of interest.
The standard model of elementary particles
is effective field theory for
weak, strong and electromagnetic interactions.
It is placed in a specific region of expansive
energy scales.
One can effectively focus on the energy scale of interest
without knowing details at other scales.
In this aspect,
the hierarchy is favorable.
On the other hand,
 the dynamical 
variables of 
the standard model possess
yet another hierarchy between parameters of
the theory itself.
It may be expected that parameters 
of a theory would be of the same order if 
the theory gives a physical 
description in a certain energy region.
For the variables of the standard model, 
fermion masses are not of the same order.

Effective field theory should have a typical energy scale which
is stabilized.
The weak scale of the standard model is not protected from
radiative corrections because
masses of fields can receive ultraviolet sensitive quantum
corrections.
Candidates to stabilize energy scales of a theory have
been presented based on
a possibility of extra dimensions~%
\cite{ArkaniHamed:1998rs,rs}.
As for 
the other hierarchy,
an explanation for generating 
fermion masses
from
order ${\cal O}(1)$ quantities was proposed in an 
extra-dimensional model
where massive fermions 
propagate in the bulk~\cite{Kaplan:2001ga}.
Here bulk masses of fermions are of the same order and
the resulting low energy fermion masses are hierarchical
via hyperbolic trigonometric functions
for overlapping wave functions, as we will
write down explicitly
in the subsequent section.
This picture is applicable for an extension of
the standard model.
It would be straightforward to construct a model,
if the values of bulk mass parameters are
chosen by hand like the standard model.
In treating extra-dimensional models, it would be important to
identify what becomes to constrain the original  
parameters in the bulk.

From a viewpoint of effective theory,
focusing on relations between values of masses is
as important as individual absolute values of the masses. 
For example, in hadron physics, the masses
of pion, kaon and $\eta$ approximately satisfy a version of the Gell-Mann--Okubo relation~%
\cite{weinberg}:
\bea
   3m_\eta^2 + 2m_{\pi^+}^2 -m_{\pi^0}^2
     =2m_{K^+}^2 +2m_{K^0}^2 .
     \label{gmo}
\eea
The $SU(3)$ properties gave insight 
to leading to quantum chromodynamics (QCD).
The pion, kaon and $\eta$
are not fundamental dynamical variables of QCD.
This implies that to find mass relations in a theory
can be a clue to discover a property in a more fundamental theory.
It is well known that charged lepton masses 
satisfy the relation~\cite{Koide}:
\bea
   m_{e} + m_{\mu} + m_{\tau}
    ={2\over 3} 
 (\sqrt{m_e} +\sqrt{m_{\mu}} + \sqrt{m_{\tau}})^2.
  \label{km}
\eea
For the central values of the pole masses
in Particle Data Group \cite{Amsler:2008zzb}:
\bea
 m_e=0.510998910, \qquad
 m_\mu = 105.658367, \qquad
 m_\tau = 1776.84 ,
  \label{pdg}
\eea
in unit of MeV,
the factor $2/3$ in Eq.~(\ref{km}) is given by
$1.999978/3$.
Mass relations for quarks can be taken into account and
they would receive large QCD corrections.
It needs to be examined if there are  
simple relations between the same-order parameters for
bulk masses similar to Eq.~(\ref{km}).

In this paper,
we consider a five-dimensional toy model for
charged leptons on the orbifold $S^1/Z_2$
where the five-dimensional spacetime is flat.
The gauge group is $SU(2)_L\times U(1)_Y$ to describe weak and electromagnetic
interactions. 
The leptons propagating in five dimensions are Dirac fermions and have
Dirac masses that are not hierarchical originally.
By orbifolding, components with zero modes which are
left-handed or right-handed survive in low energies and  
low energy theory becomes chiral.
For chiral anomaly cancellation, quarks could be introduced in a parallel way.
The low energy chiral leptons combine with each other to form 
mass terms through Yukawa couplings given on a single boundary.
The four-dimensional Yukawa couplings
are multiplied by
wave functions of the leptons and depend on their bulk masses.
The hierarchical masses are due to large suppression factors of the wave functions
depending on bulk masses.
Instead of choosing reasonable bulk masses by hand,
we study a possibility that their bulk masses 
fulfill 
simple relations like Eq.~(\ref{km}).
In other words,
we 
search for
mass relations to specify how
the 
fundamental variables of our effective theory
are constrained by substructural properties.
We find that a set of simple relations accommodates
the observed pattern of charged lepton mass spectrum.

\section{Model}

The field content is
$SU(2)_L$ gauge boson $A_M^a (x,y)$,
$U(1)_Y$ gauge boson $B_M(x,y)$,
$SU(2)_L$-doublet Higgs boson $H(x)$,
$SU(2)_L$-doublet Dirac fermion $E^i(x,y)$, 
and
$SU(2)_L$-singlet Dirac fermion $e^i(x,y)$.
The fermions are written in terms of chiral components as
\bea
   E^i (x,y) &\!\!\!=\!\!\!&
     \left(\begin{array}{c}
       E^i_L \\
       E^i_R \\
     \end{array} \right) (x,y), \qquad
   e^i (x,y) =
    \left(\begin{array}{c}
       e'{}^i_L \\
       e^i_R \\
     \end{array} \right) (x,y) ,
       \label{chlepton}
\eea
where $i=1,2,3$.
Capitalized indices
$M,N$ run over $0,1,2,3,5$,
Greek indices $\mu$ run over 0,1,2,3
and fifth index is also denoted as $y$.
We use a metric $\eta_{\mu\nu}=\textrm{diag}(1,-1,-1,-1)$.
The extra-dimensional space is compactified on $S^1/Z_2$,
where the fundamental region is $0\leq y \leq L$.
The five-dimensional spacetime is flat.
The charged leptons are included as
\bea
    E_L^i =
       \left(\begin{array}{c}
          \nu^i \\
          e^i \\
        \end{array} \right)_L
  =\left(
    \left( \begin{array}{c}
        \nu_e \\
        e^- \\
         \end{array}\right)_L ,
      \left(\begin{array}{c}
         \nu_\mu \\
          \mu^- \\
        \end{array}\right)_L ,
     \left(\begin{array}{c}
          \nu_\tau \\
            \tau^- \\
        \end{array}\right)_L \right) , 
\qquad 
  e_R^i = (e_R^- , \mu_R^-, \tau_R^-) .
    \label{compen}
\eea
In a parallel way, it is possible to include quarks  for anomaly cancellation.
We will not discuss quarks further. 
The covariant derivative is given by
\bea
   D_M  = \partial_M -ig A_M^a T^a
   -ig'{Y\over 2} B_M.
\eea
The hypercharges are assigned as $Y/2=1/2$ for the Higgs $H$, $Y/2=-1/2$ for 
the doublet $E^i$
and $Y/2=-1$ for the singlet $e^i$, which are combined with 
$T^3 =\pm 1/2$ to give the correct electric charges.

The fermion kinetic energy terms and mass terms are   
\bea
   {\cal L} = \bar{E}^i \left(i \gamma^M \partial_M
     - M_E^i \epsilon(y)\right)E^i
  +  \bar{e}^i \left(i \gamma^M \partial_M
     - M_e^i \epsilon(y)\right)e^i
  , \label{lag}
\eea
where Dirac conjugate is taken as $\bar{E}^i=E^i{}^\dag \gamma^0$ and
$\bar{e}^i =e^i{}^\dag \gamma^0$.
The bulk masses are denoted as $M_E^i$ and $M_e^i$.
For $E^i$ and $e^i$, we assume the economical case
$M_E^i = M_e^i\equiv M^i$ for each $i$.
To examine relations between masses for generations $i$
is our main analysis.
The basis for the Dirac matrices is
\begin{eqnarray}
  \gamma^M=\left(\left(\begin{array}{cc}
		 0&\sigma^\mu \\
                 \bar{\sigma}^\mu& 0\\
		       \end{array}\right),
          \left(\begin{array}{cc}
	   -i&0 \\
            0&i \\
		\end{array}\right)\right) ,
\end{eqnarray}
where $\sigma^\mu=(1,\overrightarrow{\sigma})$,
$\bar{\sigma}^\mu=(1,-\overrightarrow{\sigma})$
and the sign function is denoted as 
$\epsilon(y)=1$ for $0<y < L$ mod $2L$
and $\epsilon(y)=-1$ for $-L<y<0$ mod $2L$.

The parity assignments at $y=0,L$ are taken as
\bea
A_\mu(x,-y) &\!\!\!=\!\!\!&  A_\mu (x,y), \qquad
A_\mu(x,L-y) =  A_\mu (x, L+y) ,
\\
 A_5(x,-y) &\!\!\!=\!\!\!& - A_5 (x,y), \qquad
A_5(x,L-y) = - A_5 (x, L+y) ,
\\
B_\mu(x,-y) &\!\!\!=\!\!\!&  B_\mu (x,y), \qquad
B_\mu(x,L-y) =  B_\mu (x, L+y) ,
\\
 B_5(x,-y) &\!\!\!=\!\!\!& - B_5 (x,y), \qquad
B_5(x,L-y) = - B_5 (x, L+y) ,
\\
   E^i (x,-y) &\!\!\!=\!\!\!&
      i\gamma^5 E^i (x, y) ,
  \qquad
   E^i (x,L-y) =
      i\gamma^5 E^i (x, L+y) ,
\\
      e^i (x,-y) &\!\!\!=\!\!\!&
      -i\gamma^5 e^i (x, y) ,
  \qquad
   e^i (x,L-y) =
      -i\gamma^5 e^i (x, L+y) .
\eea
The $SU(2)_L$ gauge boson $A_\mu$ 
has the parity $+$ at $y=0$
and $+$ at $y=L$, which are denoted as $(++)$.
The parity $(++)$ are for 
the fields $A_\mu$, $B_\mu$, $E_L^i$ and $e_R^i$.
The other fields $A_5$, $B_5$, $E_R^i$ and $e_L^i$
have the parity $(--)$.
The scalars $A_5$ and $B_5$ do not develop vacuum expectation values.
The gauge group $SU(2)_L\times U(1)_Y$ is broken
to the electromagnetic $U(1)_{em}$ by a vacuum
expectation value of the
Higgs boson $H$ on a boundary.

From Eq.~(\ref{lag}),
the equations of motion are
\bea
   \left(\begin{array}{cc}
    \partial_5 -M^i \epsilon (y)
   & i\sigma \cdot \partial \\
   i\bar{\sigma}\cdot \partial &
   -\partial_5 -M^i \epsilon (y) \\
  \end{array}\right)
   \left( \begin{array}{c}
    \psi_L^i \\
    \psi_R^i \\
  \end{array}\right) = 0 .
\eea   
Here $\psi = E, e$.
This equation is written as second differential equations as
\bea
 &&  \left[\partial^2 -  (\partial_5 + M^i \epsilon)
      (\partial_5 - M^i \epsilon)\right] \psi^i_L =0 ,
\\
  &&    \left[\partial^2 -  (\partial_5 - M^i \epsilon)
      (\partial_5 + M^i \epsilon)\right] \psi^i_R =0 ,
\eea
With the mode expansion
\bea
  \psi_L^i (x,y) =\sum_{n=0}^\infty
     \psi_{L(n)}^i (x) f_{\psi^i,n} (y) ,
  \qquad
  \psi_R^i (x,y) =\sum_{n=0}^\infty
     \psi_{R(n)}^i (x) g_{\psi^i,n} (y) ,
\eea
the wave functions obey
\bea
 && (\partial_5 + M \epsilon)
      (\partial_5 - M \epsilon)f_{\psi,n} (y)
   =-M_n^2 f_{\psi,n} (y)  ,
\\
  &&    (\partial_5 - M \epsilon)
      (\partial_5 + M \epsilon)g_{\psi,n}(y) 
   =-M_n^2  g_{\psi,n}(y) ,
   \label{eigeneq}
\eea
where $M_n$ indicates the Kaluza-Klein mass.
For simplicity for reading, $i$ has been abbreviated.
These equations are written as
\bea
    (\partial_5 -M \epsilon) f_{\psi,n} (y)
    =  M_n g_{\psi,n}(y) ,\quad
  (\partial_5 +M \epsilon) g_{\psi,n} (y)
    =  - M_n f_{\psi,n}(y) ,\quad
\eea 
The solutions of mode functions are summarized in \cite{Uekusa:2008iz},
for various parity assignment.
The relevant part of our analysis is only zero mode.
For zero mode, the equations of motion
for $f_{\psi,0}$ and $g_{\psi,0}$ 
are the two independent equations,
\bea
   (\partial_5- M\epsilon) f_{\psi,0} (y)= 0,
   \qquad
   (\partial_5+ M\epsilon) g_{\psi,0} (y)= 0 .
\eea
For the parity $(++)$ for $f_{\psi,0}$,
the mode function is
\begin{eqnarray}
 f_{\psi,0}^{(++)}=
   \sqrt{M\over \sinh(L M)}
  \times
 \left\lbrace
  \begin{array}{c}
   e^{-M(|y-L|-L/2)} , ~~~~     M>0 , \\
   e^{M(|y|-L/2)} , ~~~~~~~~~  M<0 , \\
 \end{array}
 \right.
  \label{g0mp}
\end{eqnarray}
for $0\leq y\leq L$. 
The normalization constant is fixed by
$\int_0^L dy (f_{\psi,0}^{(++)})^2=1$.
For $M>0$, $\psi_L^{++}$ is localized at $y=L$
and for $M<0$, it is localized at $y=0$.
For the parity $(--)$ for $f_{\psi,0}$, $f_{\psi,0}^{(--)}=0$. 
For the parity $(++)$ for $g_{\psi,0}$,
the mode function is
\begin{eqnarray}
 g_{\psi,0}^{(++)}=
   \sqrt{M \over \sinh(LM)}
   \times 
 \left\lbrace \begin{array}{c}
e^{-M(|y|-L/2)} , ~~~~~~~     M> 0 , \\
e^{M(|y-L|-L/2)} , ~~~~~~~  M<0 , \\
\end{array}
\right.
  \label{f0mp}
\end{eqnarray}
for $0\leq y\leq L$.
The normalization constant is fixed by
$\int_0^L dy (g_{\psi,0}^{(++)})^2=1$.
For $M>0$, $\psi_R^{++}$ is localized at $y=0$
and for $M<0$, it is localized at $y=L$.
For the parity $(--)$ for $g_{\psi,0}$, $g_{\psi,0}^{(--)}=0$. 
These mean that
the fields $\psi_L^{++}$ and $\psi_R^{++}$
are localized at the opposite fixed points.
The product $f_{\psi,0}^{(++)}g_{\psi,0}^{(++)}$ is written as
\begin{eqnarray}
  f_{\psi,0}^{(++)}g_{\psi,0}^{(++)}
 ={M\over \sinh(L M)} ,
 \label{fgfixedp}
\end{eqnarray}
for arbitrary $M$ including $M=0$.

The Yukawa interaction is given by
\bea
   {\cal L}_Y = -y_{ij} \bar{E}^i \cdot
     H e^j \delta(y) + \textrm{h.c.} .
        \label{yukawa1}
\eea
If neutrinos were regarded as massless,
leptons conserve $CP$ exactly and 
the lepton number of each generation is conserved.
The charged leptons may be diagonalized.
In this case,
the Yukawa couplings can be taken as $y_{ij} =y \delta_{ij}$.
The hierarchy with respect to the generation
for the observed masses occurs from the overlapping
of wave functions.
Throughout the present section and the next section, this economical case
will be treated. 
Correspondingly to nonzero neutrino masses,
the large mixing for leptons will be taken into account in Section~\ref{sec:mixing}.
From the Yukawa interaction (\ref{yukawa1}) with $\langle H\rangle =v$ ,
the mass terms are obtained as
\bea
  {\cal L}_Y &\!\!\!=\!\!\!& - y v
  \left(  {M^1\over \sinh(L M^1)} \bar{e}_L e_R 
  +{M^2\over \sinh(L M^2)} \bar{\mu}_L \mu_R 
  +{M^3\over \sinh(L M^3)} \bar{\tau}_L \tau_R  \right) 
  +\textrm{h.c.}
\nonumber
\\
  &\!\!\!=\!\!\!& - {y \over L} v
  \left(  {c_e\over \sinh(c_e)} \bar{e}_L e_R 
  +{c_\mu \over \sinh(c_\mu)} \bar{\mu}_L \mu_R 
  +{c_\tau \over \sinh( c_\tau)} 
  \bar{\tau}_L \tau_R  \right) 
  +\textrm{h.c.}
  \label{yukawa4}
\eea
Here the dimensionless bulk mass parameters are defined as
\bea
   (c_e , c_\mu , c_\tau)
    =(L M^1 , L M^2 , L M^3) .
\eea 
The dimensionless Yukawa coupling is 
given by $Y = y M_*$ where $M_*$ denotes the fundamental
scale. 
The running of gauge coupling constants is significantly
enhanced beyond the scale of size of extra dimensions
so that $M_* L$ is not very large~\cite{M*L,enhance}.
According to \cite{Kaplan:2001ga},
the dimensionless quantities can be taken as
$Y \sim {\cal O}(1) - {\cal O}(10)$ 
and $M_* L \sim {\cal O}(10)$.
We choose $Y=1$ and $M_* L =10$. 
From Eq.~(\ref{yukawa4}), the
charged lepton masses are written in terms of
bulk mass parameters as
\bea
    m_e &\!\!\!=\!\!\!& \left({Y\over M_* L}\right)
    {c_e\over \sinh(c_e)}  \cdot v,
     \label{mece}
\\
    m_\mu &\!\!\!=\!\!\!&\left({Y\over M_* L}\right)
    {c_\mu\over \sinh(c_\mu)} \cdot v,
\\
    m_\tau &\!\!\!=\!\!\!&\left({Y\over M_* L}\right)
  \  {c_\tau\over \sinh(c_\tau)}\cdot v .
    \label{mtauctau}
\eea 
For $v$, we will adopt
$v=(2\sqrt{2}G_F)^{-1/2}=174.104$~GeV with
$G_F = 1.16637 \times 10^{-5}$~GeV${}^{-2}$.

\section{Mass relations}

We consider our five-dimensional model an effective 
theory
where hierarchical numbers can be generated
from order ${\cal O}$(1) numbers.
It still remains to be unknown what the origin of the 
masses of charged leptons is.
A clue to fundamental theory beyond
our model would be to find the existence of
simple mass relations
for the dynamical variables of our model, that is,
bulk fermions.

In this section, we 
analyze
relations between $m_e$, $m_\mu$, $m_\tau$
and between $c_e$, $c_\mu$, $c_\tau$.
The charged lepton mass relation (\ref{km})
has a geometrical expression given by~\cite{Foot:1994yn},
\bea
     \cos \theta = 
        {(\sqrt{m_e}, \sqrt{m_\mu}, \sqrt{m_\tau})
         \over
        |(\sqrt{m_e}, \sqrt{m_\mu}, \sqrt{m_\tau})|}
       \cdot \vec{n}  ,
   \label{foot}
\eea
where $\theta =45^\circ$ and the unit vector $\vec{n}=(1,1,1)/\sqrt{3}$.
We would like to look for a similar simple relation between the bulk mass parameters $c_e$, $c_\mu$ and $c_\tau$.
The mass relations required here
are equations to give a clue to 
more fundamental theory.
As a principle, they should be simple.
As the model is an effective theory,
the relations are not necessarily exact 
but can be approximate
as in Eq.~(\ref{gmo}).
A candidate of 
the value of $\theta$ may be such a rough and a 
simple value
as
$\theta=0^\circ, 30^\circ,
45^\circ, 60^\circ, 90^\circ$.
If three masses $m_e$, $m_\mu$ and $m_\tau$ were the same
value, the right-hand side of
Eq.~(\ref{foot}) would be 1.
This corresponds to $\theta =0^\circ$.
Because  $c_e$, $c_\mu$ and $c_\tau$ are less hierarchical 
than $m_e$, $m_\mu$, and $m_\tau$,
the degree $\theta$ for a mass relation like 
Eq.~(\ref{foot}) would not necessarily be
$\theta =45^\circ$.
On the other hand, the relation (\ref{foot}) includes
the unit vector $(1,1,1)/\sqrt{3}$.
For $c_e$, $c_\mu$ and $c_\tau$, we consider
four independent vectors with components being 
the same size, $\sqrt{3}\vec{n}=(1,1,1)$, 
$(-1,1,1)$, 
$(1,-1,1)$, 
$(1,1,-1)$. 
Thus we 
search for a possibility of
two mass relations for each $\theta$ and $\vec{n}$, 
\bea
     \cos 45^\circ = 
        {(\sqrt{c_e\over \sinh(c_e)}, 
   \sqrt{c_\mu\over \sinh(c_\mu)}, 
   \sqrt{c_\tau\over \sinh(c_\tau)})
          \over 
        |(\sqrt{c_e\over \sinh(c_e)}, 
    \sqrt{c_\mu\over \sinh(c_\mu)},
    \sqrt{c_\tau\over \sinh(c_\tau)})|
         } \cdot 
  {(1,1,1)\over \sqrt{3}} ,
     \label{mf1}
\eea
which is the relation (\ref{km}) with 
Eqs.(\ref{mece})-(\ref{mtauctau}) 
and
\bea
     \cos 
    \theta = 
        {(\sqrt{c_e}, \sqrt{c_\mu}, \sqrt{c_\tau})
          \over 
        |(\sqrt{c_e}, \sqrt{c_\mu}, \sqrt{c_\tau})|
         } \cdot \vec{n}.
   \label{mf2}
\eea
%
For the equations (\ref{mf1}) and (\ref{mf2}), 
the vector $(1,-1,1)$ is equivalent to $(1,1,-1)$
up to the naming of $c_\mu$ and $c_\tau$.
Therefore we can take independent unit vectors as
$\vec{n}=(n_1,1,n_3)/\sqrt{3}$ with $n_1=\pm 1$ and $n_3=\pm 1$.

A practical way to make extra-dimensional models
phenomenologically realistic is that
three charged lepton masses are an input
in obtaining three bulk mass parameters.
As seen from the equations (\ref{mece})-(\ref{mtauctau}), 
the bulk mass parameters can be regarded as
functions of the charged lepton masses as
$c_e=c_e(m_e,m_\mu,m_\tau;{Y\over M_*L}v)$,
$c_\mu=c_\mu(m_e,m_\mu,m_\tau;{Y\over M_*L}v)$,
$c_\tau=c_\tau(m_e,m_\mu,m_\tau;{Y\over M_*L}v)$.
Instead of choosing such parameters by hand, 
we 
connect
two charged lepton masses
to the two mass relations (\ref{mf1}) and
(\ref{mf2})
by employing one of the charged lepton masses 
as a dimensionful quantity.
If the observed value $m_e = 0.510998910$~MeV
in Eq.~(\ref{pdg})
is employed as
an input,
from Eq.~(\ref{mece}) the bulk mass parameter $c_e$ is
fixed as $c_e =13.7504$.
The other $c_\mu$ and $c_\tau$ are completely
solved by Eqs.~(\ref{mf1}) and (\ref{mf2}).
Thus $m_\mu$ and $m_\tau$ are derived.



For the numerical analysis, it is convenient to
rewritten the equations (\ref{mf1}) and (\ref{mf2})
as an equation for $c_\tau$ as
\bea
 && \left(\sqrt{c_\tau \over \sinh (c_\tau)}
    -2\left(\sqrt{c_e \over  \sinh (c_e)}
      +\sqrt{c_\mu \over \sinh (c_\mu)}\right)\right)^2
\nonumber
\\
   &&
 - 3 \left(
    {c_e\over \sinh (c_e)}
     +{c_\mu\over \sinh (c_\mu)}
       +4\sqrt{{c_e\over \sinh(c_e)}
          {c_\mu \over \sinh(c_\mu)}}\right) 
  =0 ,
     \label{eqctau}
\eea
with 
\bea
 \sqrt{c_\mu} = 
     -{n_1 \sqrt{c_e} +n_3 \sqrt{c_\tau}
       \over
         1-3\cos^2\theta}
       +k \,
  {\sqrt{3}\cos\theta\over
    1-3\cos^2\theta}  
\sqrt{(c_e + c_\tau)(2-3\cos^2\theta)
    +2n_1 n_3\sqrt{c_e c_\tau}} ,
   \label{withcase1}
\eea
where $k=\pm 1$.
From these equations, the solutions are obtained as in Table~\ref{tab:many}.
Here the mass parameters are shown for $c_i < 50$, 
otherwise the corresponding masses $m_i$
are clearly small.
\begin{table}[ht]
\caption{Masses in unit of MeV and bulk masses
for the relations (\ref{mf1})
and (\ref{mf2}) with a dimensionful quantity $m_e$. 
For other $k$ and $\vec{n}$, there are no solutions.
\label{tab:many}}
\begin{center}
\begin{tabular}{cccccccc}
   \hline \hline
    $\theta$ &  $k$ & $n_1$ & $n_3$ & $m_\mu$ & $m_\tau$ & $c_\mu$ & $c_\tau$\\ \hline
$30^\circ$ & $+1$ & $+1$ & $+1$&  
    17002.5 &   1149.4 &
     0.378 &  5.026  
           \\
$45^\circ$ &  $+1$ & $+1$ & $-1$ &
    1153.1  &  17056.4
   &  5.022 & 0.352 \\
$60^\circ$ &  $+1$ & $+1$ & $-1$ & 
   116.38  &  1941.1
   &  7.749 & 4.359 \\
$90^\circ$ &  $\pm 1$ & $-1$ & $+1$ &
    711.83  &  10697.9
   &  5.616 & 1.791 \\
$90^\circ$ & $\pm 1$ & $+1$ & $-1$ &  
    7.118  &  $8.124 \times 10^{-16}$
   &  10.883 & 49.098 \\   
  \hline
\end{tabular}
\end{center}
\end{table}
%
%
%
%
%
%
%
%
It is found that
for $\theta=60^\circ$, $k=+1$ and $n_1=-n_3=1$,
the values of $m_\mu$ and $m_\tau$ are
close to the observed values
$m_\mu=105.658367$~MeV and $m_\tau=1776.84$~MeV in 
Eq~(\ref{pdg}).
Therefore there exists a set of simple relations for bulk and physical masses.
The relations (\ref{mf1}) and (\ref{mf2}) with
$\theta =60^\circ$ and $\vec{n}=(1,1,-1)/\sqrt{3}$
 approximately accommodate
the observed values of charged lepton masses.

\section{Lepton flavor mixing \label{sec:mixing}}

As a large mixing for the lepton sector has been observed, 
it would be indispensable to take into account this effect.
In this section, we discuss how to include
neutrino masses and flavor mixing.

Similarly to Eq.~(\ref{chlepton}),
neutrinos are introduced as
\bea
   N^i (x,y) &\!\!\!=\!\!\!&
     \left(\begin{array}{c}
       N^i_L \\
       N^i_R \\
     \end{array} \right) (x,y), \qquad
   \nu^i (x,y) =
    \left(\begin{array}{c}
       \nu'{}^i_L \\
       \nu^i_R \\
     \end{array} \right) (x,y) ,
       \label{neutrino}
\eea
where $i=1,2,3$.
Here the neutrinos are included in
\bea
    N_L^i =
       \left(\begin{array}{c}
          \nu_N^i \\
          e_N^i \\
        \end{array} \right)_L
  =\left(
    \left( \begin{array}{c}
        \nu_{eN} \\
        e_N^- \\
         \end{array}\right)_L ,
      \left(\begin{array}{c}
         \nu_{\mu N} \\
          \mu_N^- \\
        \end{array}\right)_L ,
     \left(\begin{array}{c}
          \nu_{\tau N} \\
            \tau_N^- \\
        \end{array}\right)_L \right) , 
\quad 
  \nu_R^i = (\nu_{eR}^- , \nu_{\mu R}^-, \nu_{\tau R}^-) .
   \label{neuc}
\eea
and in Eq.~(\ref{compen}).
In order to distinguish the components of $N_L^i$ 
from the components of the lepton doublet $E_L^i$, 
the subscript $N$ has been
attached in Eq.~(\ref{neuc}).
In other words, 
the number of lepton doublets is two for each generation $i$.
The setup presented here is that
one of the linear combination is a light doublet and
the other becomes a heavy doublet with brane couplings.
Due to the linear combinations with complex numbers,
$CP$ phases are included.
The kinetic energy terms and mass terms for the fields $N^i(x,y)$ and $\nu^i(x,y)$
are  given by 
\bea
   {\cal L} = \bar{N}^i \left(i \gamma^M \partial_M
     - M_N^i \epsilon(y)\right)N^i
  +  \bar{\nu}^i \left(i \gamma^M \partial_M
     - M_\nu^i \epsilon(y)\right)\nu^i
  , 
\eea
as in the charged-lepton Lagrangian (\ref{lag}).
The small neutrino masses are achieved via an exponential suppression
for a appropriate $M_N^i = M_\nu^i$ for each $i$ .

The brane couplings to make a part of linear combinations heavy are
given by
\bea
    (\bar{E} , \bar{N}) 
    \left(\begin{array}{c}
        {\bf M}_1 \\
        {\bf M}_2 \\
        \end{array}
         \right)
         D \delta (y) + \textrm{h.c.},
         \label{branec}
\eea
with a doublet brane field $D(x)$.
Here two mass matrices are denoted as ${\bf M}_1$ and ${\bf M}_2$ and
the flavor index $i$ has been suppressed.
We define heavy doublets $H^i$ and light doublets $L^i$ as
\bea
   (\bar{H} , \bar{L}) = (\bar{E} , \bar{N}) U^{-1} ,
\qquad
  U=\left(\begin{array}{cc}
         U_1 & U_2 \\
         U_3 & U_4 \\
         \end{array} \right) .
\eea  
where $U_1,\cdots, U_4$ are unitary matrices.   
For these fields,    
the equation (\ref{branec}) is written as 
\bea
    (\bar{H} , \bar{L}) 
    \left(\begin{array}{c}
        U_1 {\bf M}_1 + U_2 {\bf M}_2 \\
        U_3 {\bf M}_1 + U_4 {\bf M}_2 \\
        \end{array}
         \right)
         D \delta (y) + \textrm{h.c.} .
\eea
The condition that $L^i$ are light yields ${\bf M_2} =-U_4^{-1}U_3 {\bf M}_1$.
Then the heavy fields have the coupling
\bea
    \bar{H}  
      (U_1 - U_2 U_4^{-1} U_3)
        {\bf M}_1 
         D \delta (y) + \textrm{h.c.} .
\eea
With the degrees of freedom for $U$, ${\bf M}_1$ and $D$,
this coupling can be diagonalized.

After the heavy fields are decoupled,
the lepton doublets are given by
$(E, N) = (U_3^{-1} L, U_4^{-1} L)$.
Then the charged lepton mass terms are given by
\bea
  {\cal L}_Y &\!\!\!=\!\!\!& - y v \sum_{i,j=1}^3
  \left(  {M^j\over \sinh(L M^j)} \bar{e}_L^i U_{3 ij} 
 e_R^j 
    \right) 
  +\textrm{h.c.} ,
\eea
instead of Eq.~(\ref{yukawa4}). 
Here $U_{3ij} = (U_3^{-1})^\dag_{ji}$
and $e_L^i$ denote the charged leptons included in the doublets $L^i$.
The neutrino mass terms are derived similarly.
Consequently the masses for charged leptons and neutrinos are obtained as
\bea
   (\textrm{diag}(m_e, m_\mu, m_\tau))_{j_1 j_2}
   &\!\!\!=\!\!\!& yv (V_{eL}^\dag)_{j_1 i} U_{3ij}
     {M^j\over \sinh(L M^j)} (V_{eR})_{jj_2} ,
\\     
     (\textrm{diag}(m_{\nu_1}, m_{\nu_2}, m_{\nu_3}))_{j_1 j_2}
  &\!\!\! =\!\!\!& yv (V_{\nu L}^\dag)_{j_1 i} U_{4ij}
     {M_\nu^j\over \sinh(L M_\nu^j)} (V_{\nu R})_{jj_2} ,
\eea
where $V_{eL}$, $V_{eR}$, $V_{\nu L}$ and 
$V_{\nu R}$ are mixing matrices for leptons.
The physical lepton mixing is given by $V_{\textrm{\scriptsize MNS}}
=V_{\nu L}^\dag V_{e L}$.
We can choose $U_3= V_{eL} V_{eR}^\dag$ with keeping 
the physical value for $V_{\textrm{\scriptsize MNS}}$.
While this choice may affect the value of $U_4$, 
the corresponding charged lepton masses are 
given in Eqs.~(\ref{mece})-(\ref{mtauctau}). 
Therefore combining this discussion 
with the result in the previous section,
a set of relations for bulk and physical masses 
accommodates the observed pattern of charged lepton masses compatibly
with the large mixing for flavor in lepton sector.
To fix all the matrices appearing here,
information on the empirical neutrino mass relation seems necessary.

\section{Conclusion}

We have studied a possibility to 
connect
masses of charged leptons 
to
 two simple mass relations.
One is the 
empirical
charged lepton mass relation 
given in Eq.~(\ref{km}).
The other is a similar relation 
imposed between bulk mass parameters being of the same order.
When the bulk mass parameters obey 
\bea        
        c_e + c_\mu + c_\tau
    = {4\over 3} 
   (\sqrt{c_e}+ \sqrt{c_\mu}- \sqrt{c_\tau})^2
        , \label{ccc}
\eea
which is the case of 
$\theta =60^\circ$ as in Eq.~(\ref{mf1}),
we have found that the observed 
pattern of charged lepton mass spectrum
is approximately reproduced.
If $Y/(M_* L)=1/11$ and
the observed $m_e$ are chosen as an input,
the other charged lepton masses are derived as
$m_\mu = 106.439$~MeV and
$m_\tau = 1788.94$~MeV, 
which 
are very close to the observed values given in (\ref{pdg}).
It has also been shown that these mass relations are viable including 
the large mixing for lepton flavors.
In addition to this conclusion,
there are several open questions as follows.

As discussed in Introduction, 
the desired thing is to induce a
more fundamental theory beyond the standard model
from a relation like Eq.~(\ref{ccc}).
Our result means that there is a set of simple 
relations for bulk masses as well as the 
observed lepton masses.
For the observed charged leptons,
the origin of the mass relation (\ref{km}) has
been studied based on family gauge symmetry
in \cite{sumino}.
As an extension of our model,
it would be useful to apply the idea of family symmetry
to our bulk mass parameter relation.

The present model is a toy model
for charged leptons.
Our procedure is applicable for models
with quarks and neutrinos.
If they are incorporated in
grand unified models,
the parameter space should be constrained further.
In such a case, our economical choices
$M_E^i=M_e^i$ and $y_{ij} =y\delta_{ij}$ 
might be fixed rather than
assumptions.
It should be clarified whether matter including
quarks and neutrinos is accommodated in
simple mass relations.

While utilizing extra dimensions for generating
hierarchical fermion masses,
we have not carefully taken into account extra-dimensional
effects to stabilize the typical scale of the model. 
In our model, the role of extra dimensions 
is regarded as an ingredient for 
a possible extension of the standard model
analogously to universal extra dimensions~%
\cite{Appelquist:2000nn}.
It needs to be examined whether
stabilization of the scale can be realized simultaneously.
In the Randall-Sundrum model~\cite{rs},
cutoff scales of theories are separated in a 
spatially  transparent way.
When the typical scale of the theory 
is stabilized in a warped spacetime,
mass relations might be found in a clearly compatible
way with the stabilization.

Finally, it would be theoretically desirable that
the radius of an extra dimension is stabilized.
A radius stabilization has been found in a model 
where the mass parameter of a bulk scalar field 
has a critical value~\cite{radius}.
Because the model is supersymmetrically constructed,
a bulk fermion as the superpartner of this scalar field
has the identical mass parameter.
When such a possibility is taken into account,
the existence of 
the critical value might add a constraint on
bulk mass parameters of fermions.

\vspace{8ex}

\subsubsection*{Acknowledgments}

I thank Yoshio Koide for valuable suggestions.
This work is supported by Scientific Grants 
from the Ministry of Education
and Science, Grant No.~20244028.

\begin{appendix}
 
\end{appendix}

\newpage



\end{document}